\documentclass[journal=nalefd,manuscript=letter,layout=onecolumn]{achemso}

\pdfoutput=1

\usepackage{array}
\usepackage{booktabs}
\usepackage{listings}
\usepackage{mathpazo}
\usepackage{microtype}
\usepackage{hyperref}

\newcommand{\beginsupplement}{%
        \setcounter{table}{0}
        \renewcommand{\thetable}{S\arabic{table}}%
        \setcounter{figure}{0}
        \renewcommand{\thefigure}{S\arabic{figure}}%
     }

\author{Suhwan Song}
\affiliation{Department of Chemistry, Yonsei University, 50 Yonsei-ro Seodaemun-gu, Seoul 03722, Korea}

\author{Min-Cheol Kim}
\affiliation{Department of Chemistry, Yonsei University, 50 Yonsei-ro Seodaemun-gu, Seoul 03722, Korea}

\author{Eunji Sim}
\affiliation{Department of Chemistry, Yonsei University, 50 Yonsei-ro Seodaemun-gu, Seoul 03722, Korea}
\email{esim@yonsei.ac.kr}

\author{Anouar Benali}
\affiliation{Argonne Leadership Computing Facility, Argonne National Laboratory, 9700 S. Cass Ave, Argonne, IL 60439, USA}

\author{Olle Heinonen}
\affiliation{Material Science Division, Argonne National Laboratory, 9700 S. Cass Ave, Argonne, IL 60439, USA}

\author{Kieron Burke}
\affiliation{Departments of Chemistry and of Physics, University of California, Irvine, CA 92697,  USA}

\title{Benchmarks and reliable DFT results for spin-crossover complexes}

\abbreviations{DC-DFT, DFT}

\begin{document}



\newpage
Density Functional Theory (DFT) calculations are used throughout nanoscience,\cite{MJG11, BSMF14, PCLS13, AACC16, ZSS16, LLAH16}   but their primary purpose is to
yield accurate relative energetics and geometries.\cite{FHGT16, GBRM16}
Spin and magnetic properties are also ubiquitous, but the reliability
of magnetic moments and other spin-dependent properties from DFT
is far less well-established.\cite{PEBS95,H04,H06,K13}
In the case of molecular transport, the well-known inaccuracies of DFT orbital
energies have been linked to the typical overestimate of calculated conductances.\cite{TFSB05, QVCL07, BAP94, R14}
Thus spintronics DFT calculations have numerous sources of uncertainty,\cite{RSH01} but for many
applications, only DFT calculations are feasible.\cite{LDSM16}

Spin-crossover complexes (SCO) provide an ideal setting for examining the reliability of
DFT for spin-dependent properties.\cite{PT04}   These have been modeled in spintronics studies, both
for the isolated molecule and coupled to the leads.\cite{M02,GBFC05,PV08,GK16,IK17}  The energy
difference between the high- (HS) and low-spin (LS) states is relatively small, and their
spin-state can be flipped by entropic effects, even at room temperature.\cite{GHS94,R09,YN10}
Because it is very difficult to isolate such complexes experimentally,\cite{MBFC11,MGC15} DFT calculations
must be benchmarked against high-accuracy quantum chemical calculations.
If DFT is unable to predict the lowest energy spin-state accurately, spintronic calculations
will certainly fail.

\def\SA{^{\rm SA}}
\begin{figure}[htb]
\includegraphics[width=0.7\columnwidth]{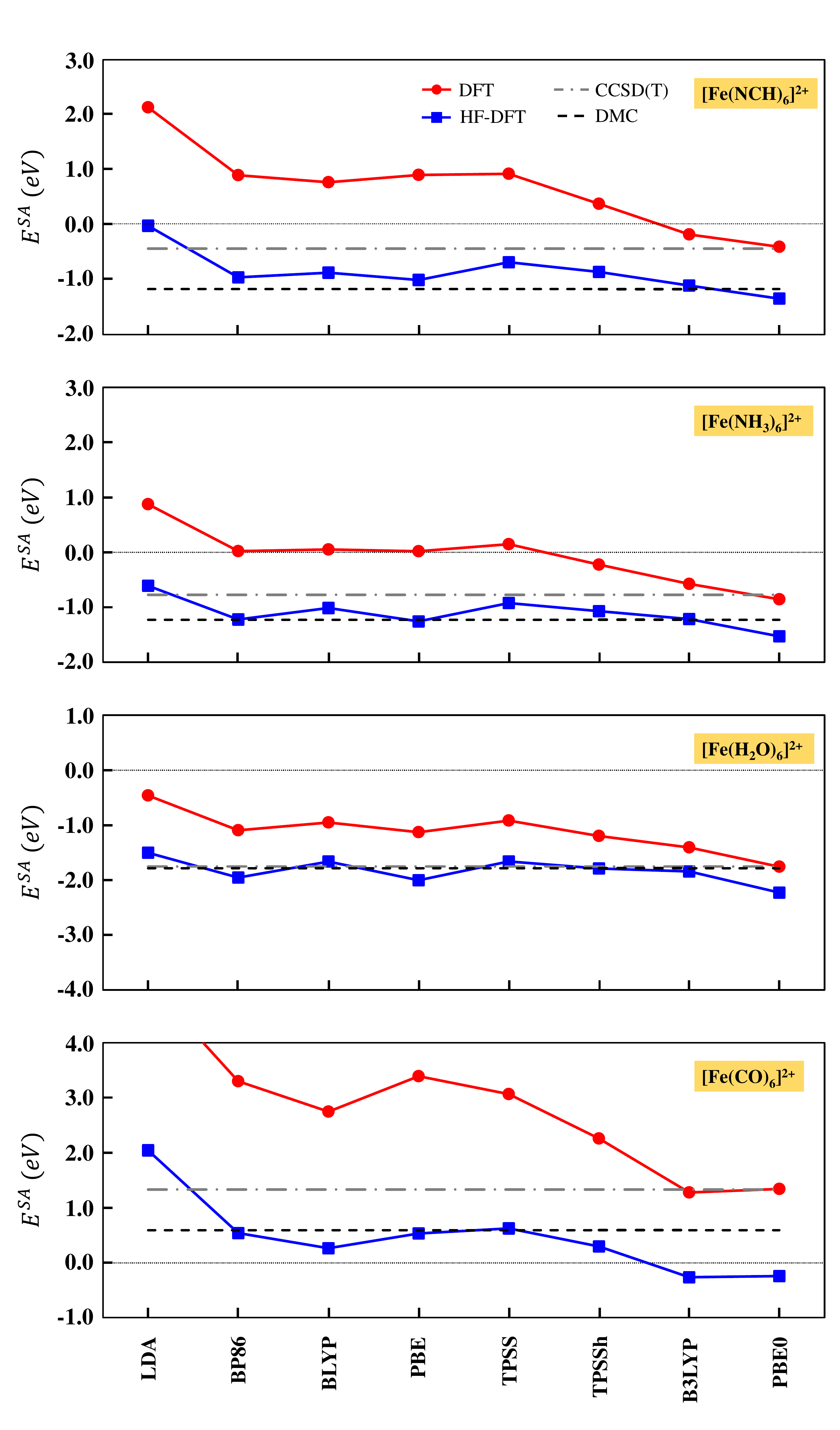}
\vskip -0.3cm
\caption{Spin adiabatic energy differences of Fe(II) complexes for
various DFT calculations and CCSD(T) (gray, unconverged) and diffusion Monte Carlo (black).}
\vskip -0.3cm
\label{fgr:SuperFigure}
\end{figure}

Here we carefully analyze DFT calculations
of the spin adiabatic 
energy difference (SA), $E^{SA}=E_{HS}-E_{LS}$, of four well-studied SCOs.\cite{FMCD04,FCDH05,DAS12,DARH12} 
These SAs are all within a few eV, due to $d$-type orbitals being close in energy.
Standard
DFT calculations give surprisingly large variations, including predicting the
incorrect ground state. 
{\em Ab initio} calculations, including CCSD(T), are notoriously difficult and fickle for 
such SCO systems.\cite{RBP10,K13}
Many of the central results of the current work are illustrated in Fig.~\ref{fgr:SuperFigure}. 
The horizontal dashed lines are extremely expensive but converged diffusion Monte
Carlo (DMC) results, showing the errors in large-basis CCSD(T) calculations which are represented as dot-dashed line.
Red points show approximate DFT results, ranging from
Localized Density Approximation (LDA) on the left to PBE0 on the right, with
increasing fraction of Hartree Fock (HF) exchange, and decreasing $E\SA$.
In particular, the CO ligand results have the largest spread, about 4 eV.
The highly accurate DMC results show that the DFT results all overstabilize the
LS state relative to HS.
The hybrids do best,\cite{FP06,DAS12,DAS13} as SA decreases with increasing fraction of exact exchange portion.
But we also include the blue points, which are the results from the various DFT
approximations evaluated on HF densities, (HF-DFT) a simple procedure suggested by the theory
of density-corrected DFT (DC-DFT).\cite{KSB11,KSB13,KSB14,KPSS15,WNJK17} 
The figures clearly show that HF-DFT is both much
more accurate and has much less variation than standard self-consistent DFT.
The rest of this paper explains how this can be.

Begin with {\em ab initio} calculations,  CCSD and
CCSD(T) on a TZVP basis yielded seemingly reasonable results, 
running about one day on a moderate
cluster.  But they have two important flaws.  When a cc-pVTZ basis was
used instead, requiring about a week to run, the results shift by up to 1.0 eV, showing
that even this basis is insufficient.  Moreover D1, a diagnostic
available in Turbomole\cite{T15}, is above 0.04 for all cases, 
whereas perturbative triples are
only reliable when D1 is order 0.02 or less,\cite{JN98}  suggesting
neither CCSD nor
CCSD(T) is reliable for
these problems.\cite{PL99} (See Table S14 of the supporting information).

What to do?
In recent decades, Quantum Monte Carlo (QMC) methods, and particularly
fixed-node diffusion Monte Carlo, (FN-DMC)\cite{WLRG01}
have proven successful at accurately describing
the properties of many solids
\cite{WLRG01,luke:solids,Santana2016,Krogel2016,Zheng2015,Schiller2015}
 and molecules.\cite{Petruzielo2012,Benali2014,Dubecky2013,Dubecky2014,Mitas:LA,Scemama2014,morales2012}
In DMC we minimize the expectation value
of the many-body Hamiltonian when propagating a
convolution of the many-body wavefunction with
the Green's function in imaginary time using many random
walkers. In the fixed node (FN) approximation, the positive and
negative regions of an initial trial wavefunction are
maintained and the walkers, which sample electrons' positions,
do not cross nodal lines. The accuracy of the wavefunction and
the obtained upper bound for the ground state energy
are limited by the quality of the nodal surface of the trial wavefunction.
A trial wavefunction  ($\Psi_T$) is often the product of a
determinant of orbitals ($\Psi_{AS}$) (fixing the nodes) and
a Jastrow function
whose parameters are found by minimizing the energy through
an initial  variational MC scheme. The accuracy of the
FN-DMC wavefunction depends solely on the quality of the nodal surface
of $\Psi_{AS}$, whose orbitals are usually taken from an effective
single-particle theory, such as HF or DFT.  
Single-determinant FN-DMC yields errors below 1~kcal/mol
for van der Waals molecules\cite{Benali2014,Dubecky2013,Dubecky2014}, 
transition metal molecules\cite{Wagner2016,Doblhoff-Dier2016}
and strongly correlated solids.\cite{benali2016,luo2016,Krogel2016,Sanata2015,Santana2016,Wagner2015,Zheng2015}   
Moreover, any such DMC result
can be systematically improved 
by increasing the complexity of $\Psi_{AS}$ by including more determinants.
Essentially exact results for the H$_2$O molecule\cite{Caffarel2016}
were obtained from a selected Configuration Interaction method, and for
the G1 test set\cite{morales2012} using a
Complete Active-Space Self-Consistent Field construct,
and is showing promising results for solids.\cite{Booth2013}
QMC also has only a weak  dependence on the basis 
set\cite{morales2012,tsatsoulis2017} and costs scale as $N^3$ with
the number of electrons $N$, allowing the study of large systems, as
the FN-DMC algorithm can efficiently use millions of processors,
including both CPUs and GPUs\cite{QMCPACK1,QMCPACK2,mathuriya2016}.

\begin{figure}[htb]
\includegraphics[width=1.0\columnwidth]{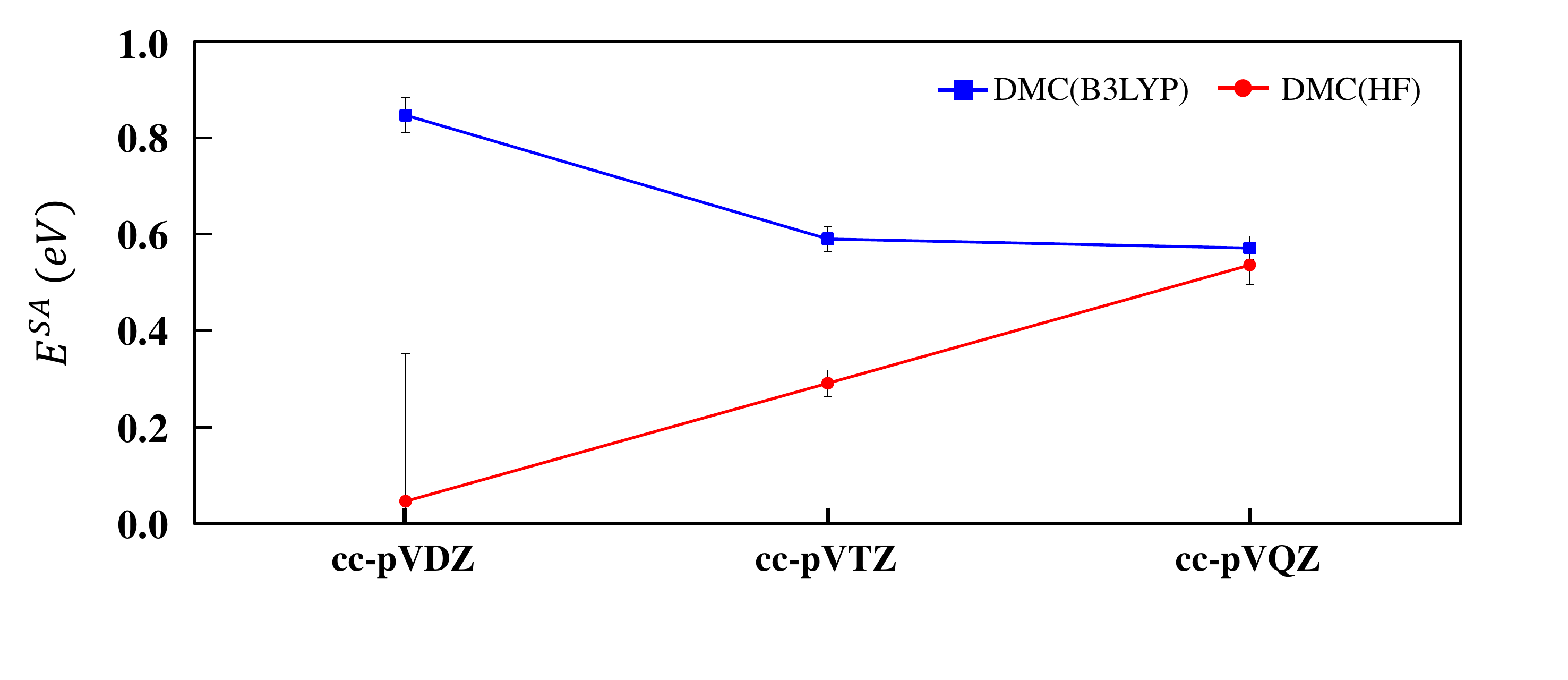}
\caption{SA of [Fe(CO)$_6$]$^{2+}$ evaluated with DMC  with two different
trial Slater determinants and three different basis sets.}
\label{fgr:Conv-DMC}
\end{figure}
As pseudopotentials for 
Fe can hold large locality errors in DMC\cite{mitavs1994quantum}, we perform all-electron
DMC.
We use a single-determinant B3LYP trial wavefunction for our DMC
calculations, and use cc-pVNZ basis sets,
where N=D,T,Q. We use the most challengling case, CO, to test convergence.
We used a time step of 0.005 and
32k walkers. To reduce the variance we used 1-, 2- and 3-body Jastrow functions
applied on all atoms and optimized for
each individual spin state. For the largest basis sets,
(cc-pVTZ and cc-pVQZ), we used 200k-400k core hours on 512
Intel Knight's Landing nodes to reach a DMC accuracy of about 25meV per compound.
Figure \ref{fgr:Conv-DMC} shows that when using a  {\em single-determinant} B3LYP trial wavefunction,
SA energies converge within 20 $\pm$ 20 meV with the size of the basis-set,
and are already converged at the cc-pVTZ level.  The HF reference recovers the
same result, but requires the larger basis set.

Despite the failure of CCSD(T), a single Slater determinant provides sufficiently
accurate nodes for DMC. Thus a system having multireference character in traditional
quantum chemical language does not necessarily imply a nodal surface that requires more
than one Slater determinant.  The 2- and 3-body Jastrow factors account for considerable
correlation.
This is consistent with claims of the weak dependence on basis-sets size in QMC.
For all remaining compounds, 
we chose to carry all calculations with all electrons, a cc-pVTZ basis set,  
a B3LYP trial wavefunction and spin-compound optimized 1-3 Body Jastrows.

In contrast with quantum chemical methods, DFT calculations converge more rapidly
with basis set. All calculations here were done in cc-pVTZ and many were compared with the Fritz Haber Institute molecular simulations package (FHI-aims), yielding agreement to within 0.01 eV in all cases.
(See Table S2 in the supporting Information.)
We next explain the difference between the
red and blue lines in Fig.~\ref{fgr:SuperFigure}.

Ground-state Kohn-Sham (KS) DFT is mostly used to extract approximate ground-state energies of electronic systems as a function of nuclear positions.\cite{KS65}  But every practical calculation uses an approximation to the unknown exchange-correlation (XC) energy, and the self-consistent cycle of a KS solver finds the density that minimizes the approximate energy for the given system.\cite{P13} Thus both the density and energy are approximated.\cite{KSB13} It is trivial to separate out the energetic consequences of these two approximations. The energy error in such a calculation may be defined as
\begin{equation}
\Delta E = \tilde E[\tilde n]-E[n],
\end{equation}
where the tilde indicates the approximation.  We define the functional ($\Delta E_F$) and density-driven ($\Delta E_D$) errors as\cite{KSB13}
\begin{equation}
\Delta E_F = \tilde E[n]-E[n],~~~\Delta E_D=\tilde E[\tilde n]-\tilde E[n],
\end{equation}
so they sum to the total energy error.  For a KS calculation, $\Delta E_F = \Delta \tilde E_{xc}[n]$, the error in the approximation to exchange-correlation, while $\Delta E_D \le 0$ by the variational principle.
In normal KS-DFT calculations, the density is so good that $|\Delta E_D| <<\Delta E_F$.
But semilocal and hybrid functionals suffer from self-interaction (or delocalization) error.\cite{CMY08}   In many cases, this error is density-driven, and $|\Delta E_D|$ becomes comparable to $\Delta E_F$.
We call such calculations {\em abnormal}, and the error can be significantly reduced by using a more accurate density.  For atoms and molecules, the HF density is often sufficient, so that HF-DFT yields much better results.  This has been shown to reduce errors for anions,\cite{KSB11} transition state barriers,\cite{VPB12} ions and radicals in solution,\cite{KSB14} and dissociating heterogeneous diatomics,\cite{KPSS15} as discussed in 
a recent review.\cite{WNJK17}

\begin{figure}[htb]
\includegraphics[width=1.0\columnwidth]{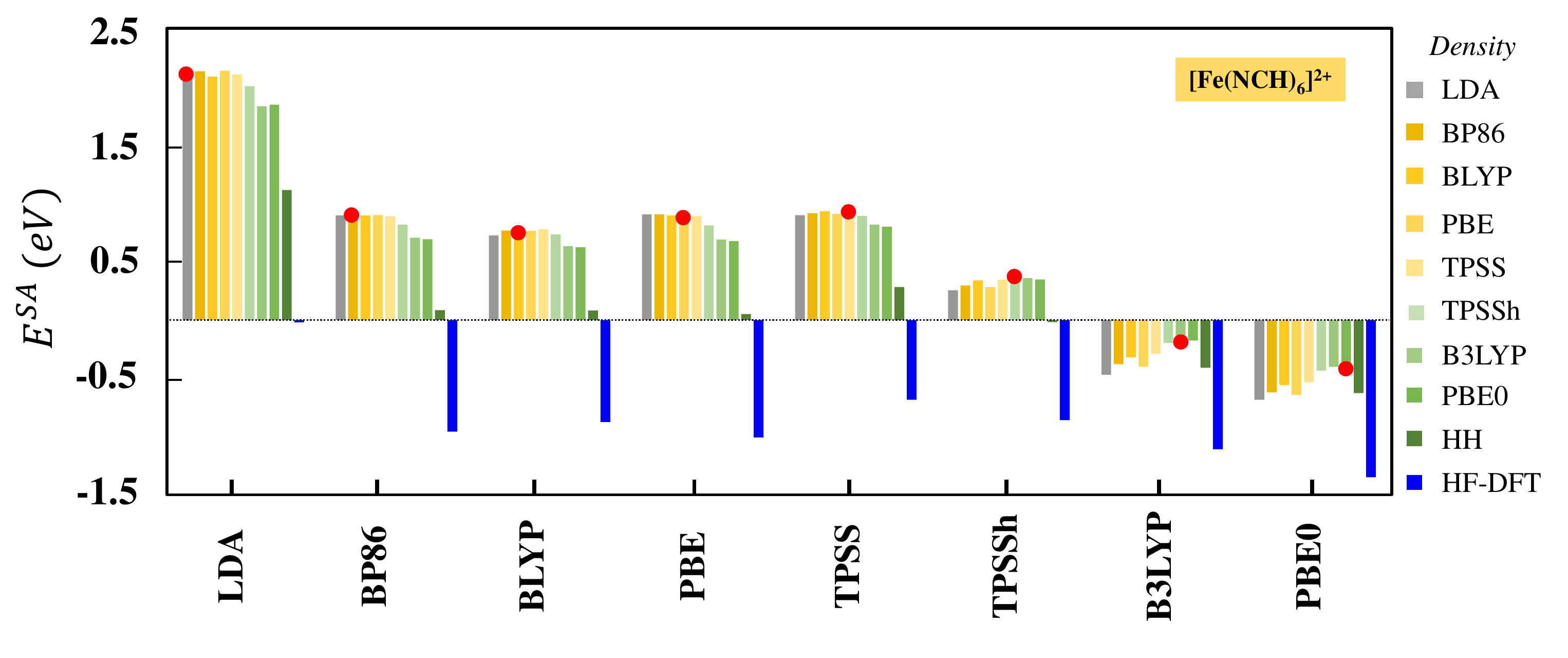}
\caption{Spin adiabatic energy difference of [Fe(NCH)$_6$]$^{2+}$ evaluated with different DFT approximations 
on several different self-consistent (red dots) and HF (blue)
densities.   (
Dark green bar is the self-consistent density of Becke's half-and-half functional (HH),
with 50\% exact exchange.)}
\label{fgr:rainbow}
\end{figure}

To demonstrate 
sensitivity to the density, in Fig.~\ref{fgr:rainbow} we show SA for several different
approximations on different densities.  
We consider here simple, commonly-used DFT approximations. 
LDA uses only the density at a point, 
generalized gradient approximations (GGAs) use both the density and its gradient, 
while (global) hybrids mix some exact exchange with GGA\cite{becke1993}:
\begin{equation}
E_{xc}^{hyb}[n]= E_{xc}^{GGA}[n]+a (E_{x}^{HF}[n]-E_{x}^{GGA}[n]),
\end{equation}
where $a$ is the amount of exchange mixing.  
In addition, we include metaGGAs (mGGAs), which also depend on a local kinetic
energy density, in the generic term GGA.
In practice, all approximations are spin-density
functionals, so $n(\textbf{r})$ represents spin-densities here.
For a given energy approximation,
there is little difference between evaluation on LDA, GGA or mGGA densities.
However, the result changes with the amount of exchange mixing.  TPSSh has 10\% and
B3LYP and PBE0 have 20\% and 25\%, respectively, while Becke's half-and-half (HH) has 50\%.  The SA
decreases with increasing $a$ in the density and even becomes correctly
negative with a HF density (100\% exact exchange).  Similar results are found for all
four complexes.  The strong dependence on
$a$ for the density demonstrates that these DFT calculations are {\em abnormal}, i.e., a significant
fraction of the error in such calculations is density-driven.
We recommend the use of such 2D {\em rainbow} plots, Fig.~\ref{fgr:rainbow}, as a check for density-driven errors,
as they do not require knowledge of the accurate result.

\begin{table}
\caption{Errors in SA in eV relative to DMC. Note that DMC values in eV are -1.17 for NCH, -1.23 for NH$_3$, -1.78 for H$_2$O and 0.59 for CO.  
The column MAE is mean absolute errors over all 4 molecules.  Rows 
AVG average over all GGA and hybrid energies, while RMSD gives 
their root-mean square deviation.
}
\begin{tabular}{lccccc}
\hline
Method&NCH&NH$_3$&H$_2$O&CO&MAE \\
\hline
\hline
&&\textbf{Ab initio}&&& \\
HF	&-2.45	&-2.04	&-1.79	&-5.24	&2.88	\\
CCSD	&0.28	&0.13	&-0.19	&-0.20	&0.20\\
CCSD(T)	&0.74		&0.45		&0.03	&0.74		&0.49\\
\hline
&&\textbf{SC-DFT}&&& \\
SVWN	&3.31	&2.10	&1.32	&4.51	&2.81	\\
&&\textit{GGA}&&& \\
BP86	&2.07	&1.25	&0.69	&2.71	&1.68	\\
BLYP	&1.94	&1.28	&0.83	&2.16	&1.55	\\
PBE	&2.08	&1.25	&0.65	&2.80	&1.69	\\
TPSS	&2.10	&1.37	&0.86	&2.48	&1.70	\\
&&\textit{Hybrid}&&& \\
TPSSh	&1.55	&1.00	&0.58	&1.67	&1.20	\\
B3LYP	&1.00	&0.65	&0.37	&0.68	&0.68	\\
PBE0	&0.77	&0.37	&0.02	&0.75	&0.48	\\
\hline
AVG	&1.64	&1.02	&0.57	&1.89	&1.28	\\
RMSD	&0.52	&0.35	&0.27	&0.82	&0.49	\\
\hline
&&\textbf{HF-DFT}&&& \\
SVWN	&1.15	&0.62	& 0.28	&1.45	&0.88	\\
&&\textit{GGA}&&& \\
BP86	& 0.21	& 0.00 & -0.17 	&-0.05	&0.11 	\\
BLYP	& 0.30	& 0.21 & 0.12 	&-0.33	&0.24 	\\
PBE	& 0.17	&-0.03 & -0.22 	&-0.06	&0.12 	\\
TPSS	& 0.49	& 0.30 & 0.12 	& 0.03	&0.24 	\\
&&\textit{Hybrid}&&& \\
TPSSh	& 0.31	& 0.15 & -0.01 	&-0.30	&0.19 	\\
B3LYP	& 0.07	& 0.01 & -0.06 	&-0.86	&0.25 	\\
PBE0	&-0.17	&-0.30 &-0.45 	&-0.84	&0.44 	\\
\hline
AVG	& 0.25	& 0.15 &0.16 	& 0.35	&0.23 	\\
RMSD	& 0.19	& 0.18 &0.19 	& 0.34	&0.23 	\\
\hline
\hline
\end{tabular}
\label{tbl:energy}
\end{table}

Knowing that there is a strong-density-driven error
does not guarantee that a HF density is sufficiently close to
the exact density to improve the DFT results, but
Fig.~\ref{fgr:SuperFigure} strongly suggests that this is the case here.  
Table~\ref{tbl:energy} reports the errors in that figure.
(Results of cc-pVTZ can be found in Tables S1-S4 of the supporting information and Tables S11 and S12 are those of TZVP.)
The first few rows show the results of {\em ab initio} quantum chemical methods.
HF is very poor, with typical errors of about 3~eV.
CCSD does much better,
but inclusion of perturbative triples doubles the errors, unlike in main-group
chemistry, suggesting
a multi-reference nature, corroborated by the high D1 values.
To check the reliability of our DMC with just a single Slater determinant trial wavefunction, for the CO complex,
we constructed a multi-determinant trial wavefunction consisting of 135~k determinants  
using the perturbatively selected configuration interaction method (CIPSI) as described in
Ref.~\citenum{giner2013}.  In order to make the calculation tractable, and only for this test, we used a pseudopotential for the Fe atom.
We compared the variational energies from this multideterminant expansion with the energies obtained from DMC with the B3LYP nodal surface using  the same pseudopotential and basis set. The calculations yielded the same $E^{SA}$ to within a $\pm0.005$~a.u. error bar.
More details of this calculation will appear elsewhere.

Another signal that a self-consistent DFT calculation might suffer
from a density-driven error is an abnormally small HOMO-LUMO gap,
often less than 1~eV.\cite{KSB13}  
The gaps for all molecules and functionals are listed
in the supporting Information, talbe S10, for both HS and LS states.  
For the HS states, all gaps are below 1~eV,
with some as small as 0.4~eV.

In self-consistent DFT, typical LDA errors are similar in magnitude but opposite in sign to those of HF. 
All DFT approximations overstabilize the LS state relative to the HS state in a very systematic way.  
Typical GGAs are about 1.7 eV too high. 
The best functional is the hybrid PBE0 with a mean absolute error of 0.5 eV, 
and the performance of other hybrids depend on the amount of mixing. 
Averaging over all 4 GGAs and 3 hybrids yields 1.3 eV error, 
and the root-mean-square deviation (RMSD) from this value is about 0.5 eV.  
But when evaluated on HF densities, the typical LDA error is reduced by a factor of 3, 
while GGA errors are below 0.25 eV and are now comparable to hybrids. 
The average HF-DFT error is now within 0.25 eV, with RMSD deviations of 0.23 eV.
Clearly, both the accuracy of the
typical functional is substantially higher, and the variations
among functionals much lower than in self-consistent DFT.
Moreover, the most accurate results are from either BP86 or PBE, both simple GGAs,
with no exchange mixing.  BP86 was previously identified as (at the 
time of testing) the most accurate for transition metal complexes.\cite{FP06} 
Thus HF-DFT yields much more reliable and accurate estimates of SA for these molecules.

\begin{figure}[htb]
\includegraphics[width=0.9\columnwidth]{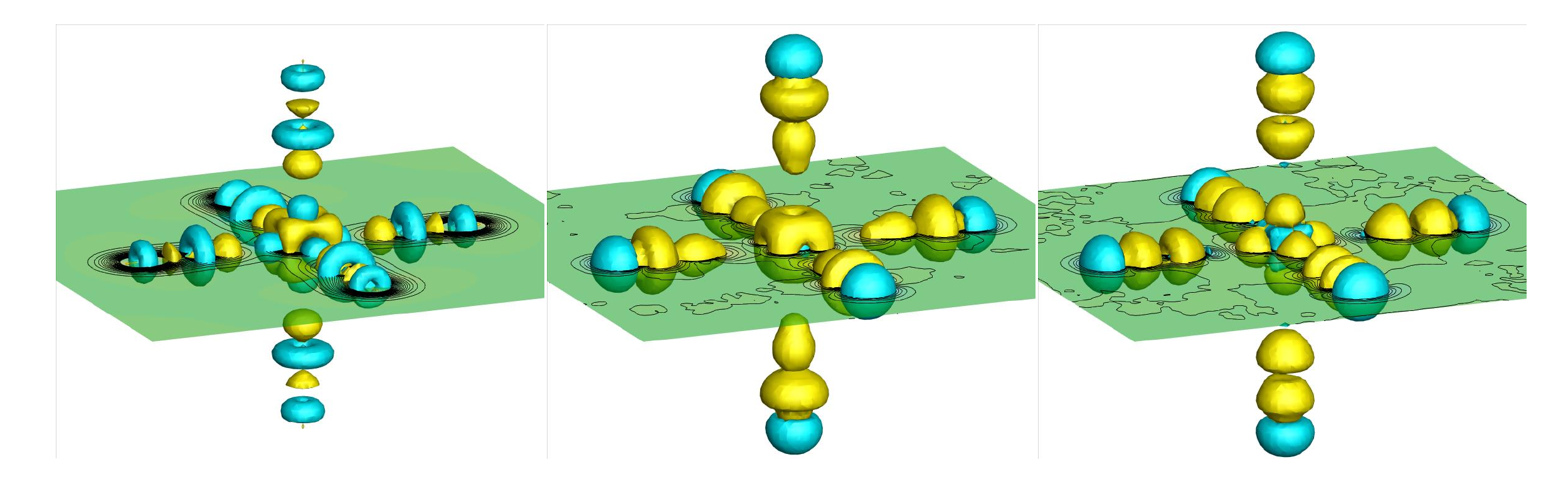}
\caption{Isosurface plots of difference densities at $\pm0.1$~{\AA}$^{-3}$ for the LS [Fe(CO)$_6$]$^{2+}$ complex, with gold (blue) +0.1 (-0.1)~{\AA}$^{-3}$. Left panel: $n^{HF}-n^{PBE}$. Middle panel: $n^{DMC}-n^{PBE}$. Right panel: $n^{DMC}-n^{HF}$.}
\label{fgr:LS_CO_QMC-DFT-PBE_3D}
\end{figure}

To relate density differences to density-driven errors, we focus on 
LS [Fe(CO)$_6$]$^{2+}$ as an example, comparing densities using DMC with the GAMESS B3LYP
all electron trial wavefunction with PBE and HF densities obtained with FHI-aims. 
The left panel of Fig. \ref{fgr:LS_CO_QMC-DFT-PBE_3D} shows isosurfaces of the HF and PBE electron number density difference
at $\pm 0.1$~{\AA}$^{-3}$, with gold (blue) positive (negative) values.
The HF density polarizes the oxygen more, shifting density towards the Fe ion at the center. This is visible as gold (positive density difference) regions closer to the Fe ion. Also, the PBE density shows much more pronounced lobes from the $3d_{z^2}$ and the $3d_{x^2-y^2}$ orbitals, visible as blue (negative density difference) lobes near the center. The middle panel shows the DMC and PBE density difference. This is {\em qualitatively} similar to the HF and PBE density difference in that the oxygen is more polarized in DMC than PBE, and also by the less pronounced $3d$ orbital lobes at Fe ion (visible as dimples with blue centers in the gold cuboid). Finally, the right panel shows the DMC and HF density difference. 
This figure shows that DMC polarizes the oxygen more than does HF, and also that the Fe $3d$ lobes are somewhat more extended in DMC than HF. We speculate that the key reason the HF-DFT scheme yields quantitatively more accurate energies than DFT is that the HF Fe $3d$ lobes, like in DMC, are less pronounced than in DFT.

\begin{figure}[htb]
\includegraphics[width=0.7\columnwidth]{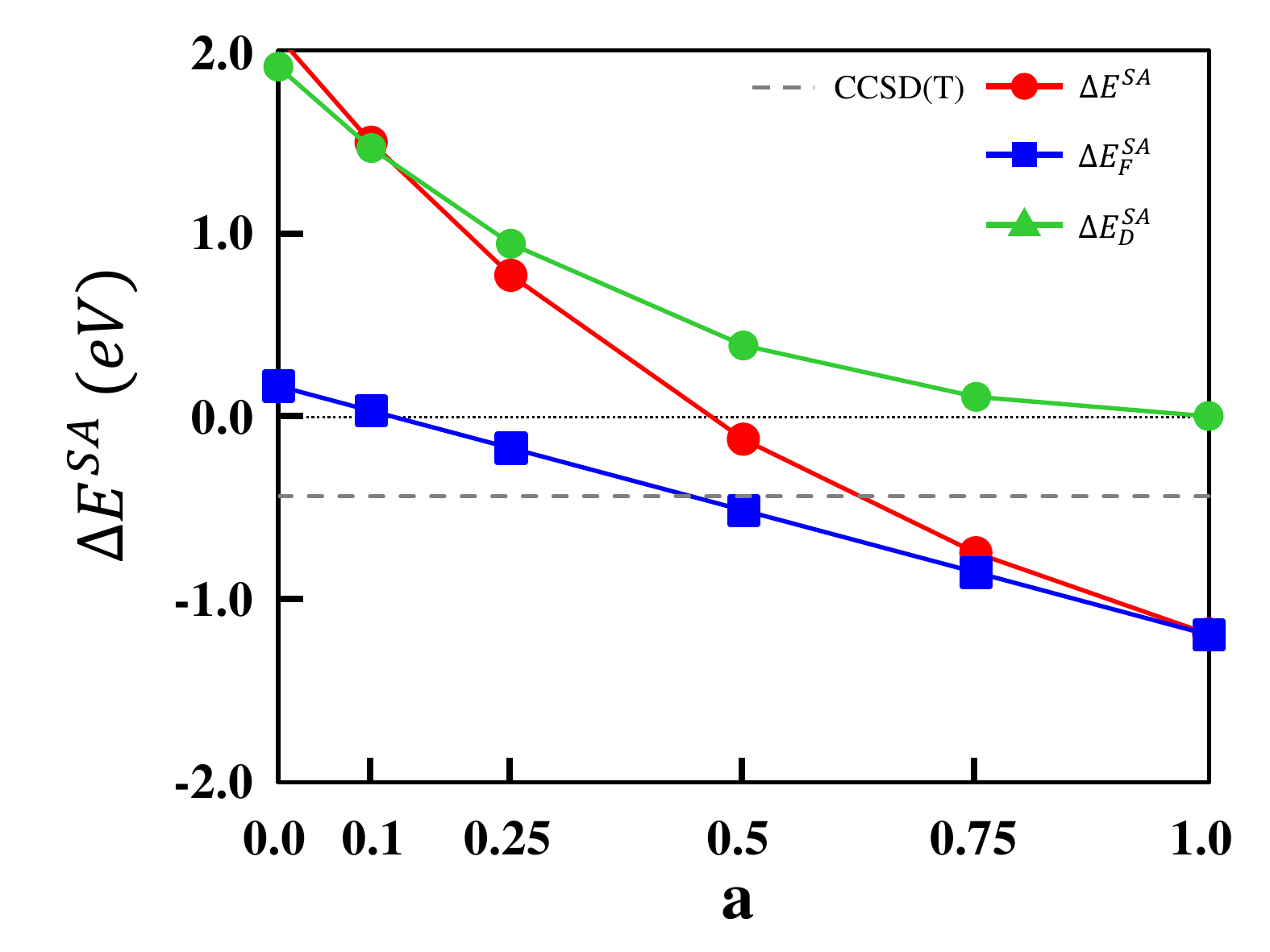}
\caption{Error in spin adiabatic energy difference of [Fe(NCH)$_6$]$^{2+}$ complex from $a$PBE, decomposed into functional and density-driven components.}
\label{fgr:dE}
\end{figure}

Now, we analyze the effect of the exchange mixing parameter $a$ for NCH, assuming that the HF density is exact.
To do this, we run both self-consistent and HF-DFT
calculations for $a$PBE, which is the PBE functional with a variable fraction $a$ of exact exchange. 
If we assume that the self-consistent density of $1.0$PBE (i.e., 100\% exact exchange) is negligibly different
from the exact density, all components are easy to calculate within HF-DFT, and
are plotted in Fig. \ref{fgr:dE}.
The functional error is almost perfectly linear, reflecting the linear dependence on the exact
exchange energy.  The density-driven error vanishes for $a=1$, based on our assumptions. 
The plot shows that, for the pure GGA ($a=0$), the error is almost
entirely density-driven.  Turning on $a$ reduces the density-driven error,
but increases the functional error (the parameter dilemma).  
For some intermediate value of $a$, here about 0.45, the two contributions cancel,
but there's no reason that this amount of mixing won't change from molecule to molecule.  
The parameter dilemma most often arises when the position of the HOMO or
the size of the gap is important for some property or prediction\cite{}.  Then
often increasing $a$ improves the positions of the orbitals or the gap,
but worsens the underlying energetics.  HF-DFT avoids
this Procrustean dilemma, as the HF density is typically very similar to a KS
calculation with the exact exchange, whose potential has excellent orbital properties.

\begin{figure}[htb]
\includegraphics[width=1.0\columnwidth]{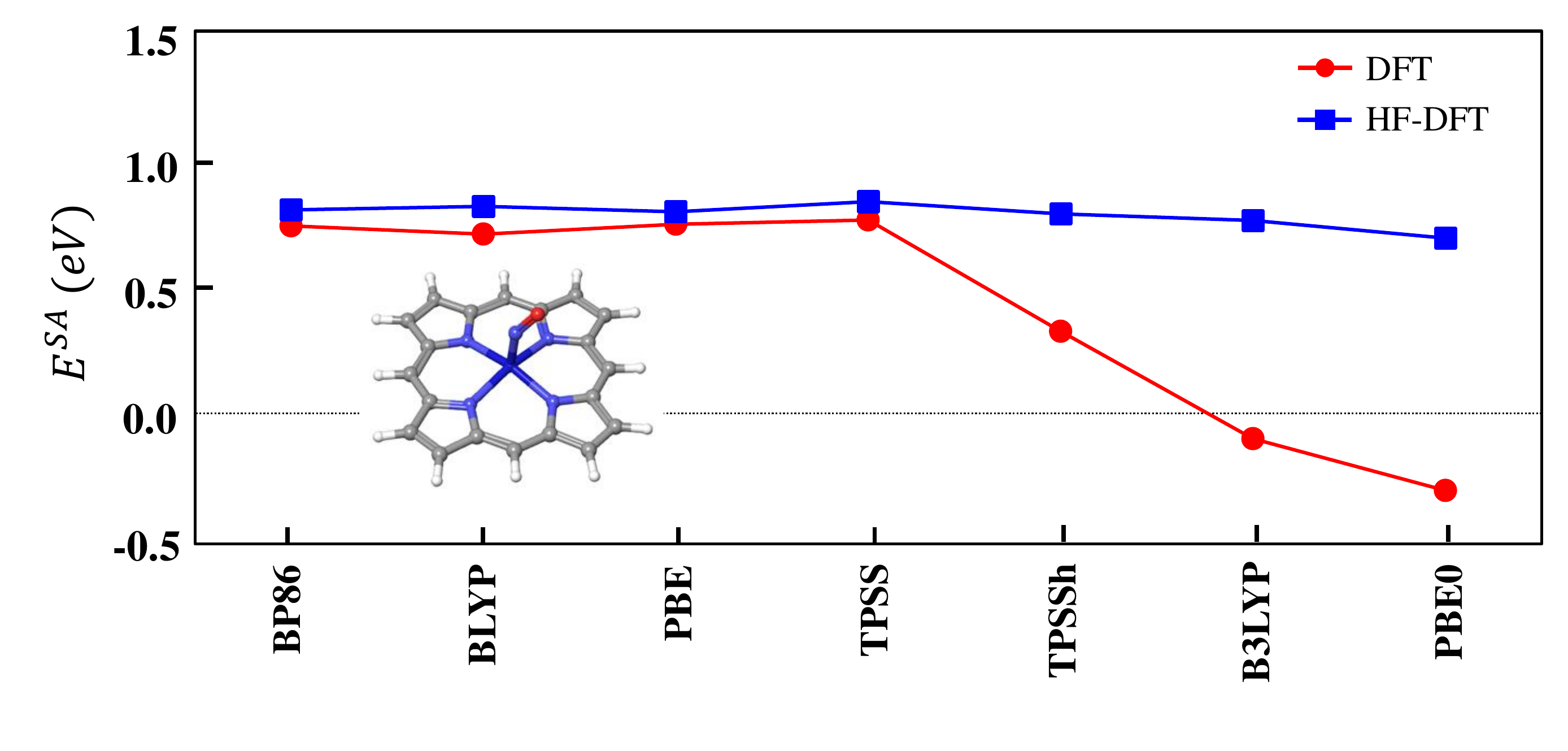}
\caption{Spin adiabatic energy difference of Fe(P)NO where P denotes porphyrin.}
\label{fgr:fepno}
\end{figure}

Lastly, we show
DFT calculations on the Fe-Porphyrin complex.  This is too large to perform either 
CCSD(T) or DMC in the basis sets used throughout this paper.  But Fig.~\ref{fgr:fepno}
shows the results of self-consistent and HF-DFT calculations.  Fig.~\ref{fgr:fepno}
clearly demonstrates the tremendous reduction in variation among approximate
energies when the HF density is consistently used, and the unambiguous
prediction that 
the LS state is lower by about 0.8 eV, 
in contradiction to self-consistent B3LYP or PBE0. 
Fe(P)NO complex is known to be strongly correlated\cite{RBP10} and, furthermore, requires a multi-reference treatment for accurate description.\cite{BJR11,BMLR12}
Nevertheless, HF-DFT successfully eliminates the chronic problem of its variational counterpart, the functional dependency.
 
We have shown that standard quantum chemical methods fail for the SA of Fe(II) SCOs. 
Semilocal DFT calculations give wildly different
results and suffer from the parameter dilemma, while
even with a decent basis set, CCSD(T) calculations are
both unconverged and unreliable.
But extremely demanding DMC methods can be converged to provide
benchmarks for these systems.  All-electron calculations using up to 3-body
Jastrow factors converge in cc-pVQZ.  Surprisingly, the B3LYP single Slater determinant
is a sufficiently accurate trial wavefunction, and converges more rapidly
than the HF wavefunction.

We have also shown that
HF-DFT yields much better
results (errors reduced by a factor of 3 or more) for these calculations, and do
not suffer from a parameter dilemma.
This is because these calculations are {\em abnormal}\cite{WNJK17}: Semilocal DFT calculations
of the SA of these compounds are contaminated by
density-driven errors.  Their small HOMO-LUMO gap for the HS state is consistent with this
hypothesis.  In such cases, and when a HF calculation does not suffer from 
spin-contamination, often using a HF density greatly reduces the DFT error, as is
the case here.  This does not mean that the HF density is generally `better' than
the self-consistent DFT density.  It simply means that the semilocal DFT SA is 
more accurate on the HF density, as it does not suffer from the
delocalization error in the density of semilocal DFT.
This is a property of the type of DFT approximation, the specific energy calculated,
and the specific system under study\cite{WNJK17}.  
We caution against overgeneralization of these results.

\section {Methods}  
All HF, CCSD(T) and DFT calculations presented in 
Figs.~\ref{fgr:SuperFigure} and \ref{fgr:rainbow} as well as in 
Table~\ref{tbl:energy} were carried out with TURBOMOLE v7.0.2\cite{T15} and 
LDA (SVWN5), GGA (PBE, BP86, BLYP), mGGA (TPSS), and
hybrid (TPSSh, B3LYP, PBE0) functionals were used for the DFT and HF-DFT calculations.
The scripts for performing HF-DFT energy calculations are available.\cite{TCCL}
The geometries used were B3LYP with TZVP basis set.  
Geometry information can be found in Tables S5-S8 of the supporting information.
We also performed fixed-spin HF and DFT calculations using FHI-aims 160328\_3.\cite{FHI-aims}
All FN-DMC (QMC) calculations used the QMCPACK code\cite{QMCPACK1,QMCPACK2,mathuriya2016}.
Relativistic corrections were ignored as Fe is rather light and relativistic effects are usually not important. Adding zero order scalar relativistic corrections within FHI-aims (control keywords "relativistic atomic zora") for CO and H$_2$O changed the $E_{SA}$ results by at most 0.2 eV, worsening agreement with DMC.

\begin{acknowledgement}

This work at Yonsei University (2016-22-0121) was supported by the grants from the Korean Research Foundation (2017R1A2B2003552, 2014R1A1A3049671) and the Ministry of Trade, Industry \& Energy, Rep. of Korea (No.10062161).
K.B. acknowledges DOE grant DE-FG02-08ER46496.
AB and OH were supported by the U.S. Department of Energy, Office of Science, Basic Energy Sciences, Materials Sciences and Engineering Division, as part of the Computational Materials Sciences Program and Center for Predictive Simulation of Functional Materials. Diffusion Monte Carlo calculations were carried through an award of computer time provided by the Innovative and Novel Computational Impact on Theory and Experiment (INCITE) program. This research has used resources of the Argonne Leadership Computing Facility, which is a DOE Office of Science User Facility supported under Contract DE-AC02-06CH11357.
We thank to Professor Frederick R. Manby and Professor Filipp Furche for stimulating discussions.

\end{acknowledgement}

\begin{suppinfo}

The following files are available free of charge.

\begin{itemize}
\item Computational details
\item Master data with cc-pVTZ basis set
\item Data with TZVP basis set 
\end{itemize}

\end{suppinfo}
\clearpage
\providecommand{\latin}[1]{#1}
\makeatletter
\providecommand{\doi}
  {\begingroup\let\do\@makeother\dospecials
  \catcode`\{=1 \catcode`\}=2 \doi@aux}
\providecommand{\doi@aux}[1]{\endgroup\texttt{#1}}
\makeatother
\providecommand*\mcitethebibliography{\thebibliography}
\csname @ifundefined\endcsname{endmcitethebibliography}
  {\let\endmcitethebibliography\endthebibliography}{}

\clearpage
\beginsupplement


\section{Supporting Information} 

All HF, DFT, HF-DFT, and CCSD(T) calculations were carried out with TURBOMOLE v7.0.2.\cite{T15S} We used the LDA, GGA (PBE, BP86, BLYP), mGGA (TPSS), Hybrid (TPSSh, B3LYP, PBE0, HH) functions to investigate the functional dependency across the various stages of "Jacob's ladder".\cite{P13S} When optimizing the structure using LDA, GGA, and mGGA functions, it has well-known energetic issues, such as the LS over-stabilization and/or the over-binding molecular bonding.\cite{DAS12S,K13S} For instance, in the HS spin state of octahedral geometry, it is experimentally well known that an axial metal-ligand bond distance is 10-20\% larger than that of an equatorial bond, but GGA often predicts all metal-ligand bonds to be equivalent. Such a characteristics is, in fact, a representative feature of the LS structure. Therefore, for each spin state of the [Fe(L)$_6$]$^{2+}$ complex, we used the optimized structure using B3LYP, regardless of the method used for energy calculations. In order to reduce the self-interaction error while maintaining the DFT level of computational efficiency, the corrected density was obtained from the HF calculation and the improved HF-DFT energy was obtained by applying the HF density to the approximate density functional. HF often results in a poor density when small basis set is used,\cite{NHHZ07} thus, we employed cc-pVTZ basis set\cite{D89,BP05} for all HF, DFT, HF-DFT, and CCSD(T) calculations. 
For all HF and DFT caculation, we adopted 10$^{-8}$ self-consistent energy and 10$^{-6}$ for one-electron density matrix convergence criteria.
Lebedev's spherical grid integration was used for the DFT spherical integration and gridsize 6, i.e. 974 gridpoints were used.
Multipole accelerated RI-J was used for the fast evaluation of the Coulomb potential.\cite{TA95}
All calculations on open-shell complexes were performed in C$_1$ symmetry.

We found agreement for self-consistent DFT and HF
to within less than 0.1~eV on all SA energies when cc-pVTZ
basis was used in TURBOMOLE and compared with the
"really tight" default basis set and mesh in FHI-aims. The
HF and PBE densities in Fig. 3 were generated using the
default "really tight" setting for meshes and basis sets.
Total electronic densities were collected from these calculations for comparison with DMC and GAMESS densities on a 10~{\AA}$^3$ Cartesian grid with a spacing of 0.1~{\AA} along the $x$- $y$, and $z$-directions.

A three-body Jastrow function reduced significantly the variance of the systems, and
its
parameters were carefully converged within variational MC. 
Both HF and DFT $\Psi_{AS}$ were used. They were generated and converged
using the GAMESS package\cite{GAMESS1,GAMESS2} with a Gaussian basis set
and an Effective Core Potential (ECP) developed specifically
for many-body theory by Burkatzki, Filippi and Dolg\cite{burkatzki2007,burkatzki2008}.
The ECP introduces errors below the FN-DMC error bar
while reducing significantly the computation time. 
For each atom, we use the basis set for which the ECP was developed: 
ANO-ccpVTZ\cite{burkatzki2008} for Fe and a cc-pVTZ\cite{burkatzki2007}
for the remaining atoms, respectively.  (Increasing to quintuple zeta
produced changes smaller than the error bars
of FN-DMC, confirming previous observations of weak basis-set dependence.
Initial  $\Psi_{AS}$(DFT) wavefunctions were generated using 
B3LYP\cite{becke1993,Stephens1994}, as its energies
were about 
0.13$\pm$0.03 eV below those from LDA wavefunction\cite{perdew1981}.
To reach a statistical error bar below 0.03 eV per molecule
and to avoid any correlated sampling or population bias,
we used 8192 walkers.  A time step of 0.001 a.u. introduced negligible error.

\newpage
\section{Raw data with cc-pVTZ basis set}

\begin{table}[]
\centering
\caption{Raw data of [Fe(L)$_6$]$^{2+}$ where L is CO, and H$_2$O. HF-DFT indicates DC-DFT with HF density.\cite{TCCLS}}
\label{my-label}
\begin{tabular}{lcccc}
\hline
 & \multicolumn{2}{c}{CO} & \multicolumn{2}{c}{H$_2$O} \\
 & HS & LS & HS & LS \\ \hline
\multicolumn{5}{c}{AB INITIO} \\ \hline
HF & -1938.596688 & -1938.425659 & -1718.487158 & -1718.355787 \\
CCSD & -1941.486969 & -1941.501497 &-1720.712183 &-1720.639633  \\
CCSD(T) & -1941.61279 & -1941.661758 &-1720.774355 &-1720.7099 \\ \hline
\multicolumn{5}{c}{SC-DFT} \\ \hline
SVWN & -1935.562476 & -1935.75007 & -1716.324658 & -1716.307699 \\
BP86 & -1943.524592 & -1943.645874 & -1722.303555 & -1722.263269 \\
BLYP & -1943.282952 & -1943.383951 & -1722.036727 & -1722.001664 \\
PBE & -1942.315225 & -1942.439855 & -1721.33526 & -1721.293755 \\
TPSS & -1943.431767 & -1943.544462 & -1722.135513 & -1722.101699 \\
TPSSh & -1943.318874 & -1943.401835 & -1722.071487 & -1722.027418 \\
B3LYP & -1942.918583 & -1942.965452 & -1721.807611 & -1721.755856 \\
PBE0 & -1942.307023 & -1942.356275 & -1721.373081 & -1721.308412 \\
HH & -1942.925932 & -1942.898408 & -1721.836892 & -1721.764887 \\ \hline
\multicolumn{5}{c}{HF-DFT} \\ \hline
SVWN(HF) & -1935.324969 & -1935.400057 & -1716.179002 & -1716.123761 \\
BP86(HF) & -1943.359619 & -1943.379374 & -1722.205822 & -1722.133963 \\
BLYP(HF) & -1943.098747 & -1943.108379 & -1721.922532 & -1721.861337 \\
PBE(HF) & -1942.149806 & -1942.169324 & -1721.240425 & -1721.166759 \\
TPSS(HF) & -1943.300547 & -1943.323445 & -1722.05761 & -1721.996494 \\
TPSSh(HF) & -1943.216065 & -1943.226898 & -1722.011766 & -1721.945904 \\
B3LYP(HF) & -1942.803829 & -1942.7941 & -1721.737777 & -1721.670029 \\
PBE0(HF) & -1942.221245 & -1942.212253 & -1721.326427 & -1721.244492 \\
HH(HF) & -1942.88147 & -1942.835129 & -1721.810121 & -1721.732972
\end{tabular}
\end{table}

\begin{table}[]
\centering
\caption{Raw data of [Fe(L)$_6$]$^{2+}$ where L is CO and H$_2$O. "aimsRT2" indicates FHI-aims\cite{FHI-domain} with the default "really tight" settings for basis set, mesh, and angular momentum expansion of the Hartree potential; "aimsRT3+2" indicates the "really tight" default setting for mesh and expansion of the Hartree potential and the Tier 3 basis set with added two Tier 4 basis functions for Fe (hydro 5 f 12 and hydro g 10.4). Relativistic corrections were turned off with the setting "override\_relativity .true."}
\label{my-label}
\begin{tabular}{lcccc}
\hline
 & \multicolumn{2}{c}{CO} & \multicolumn{2}{c}{H$_2$O} \\
 & HS & LS & HS & LS \\ \hline
\multicolumn{1}{c}{} & \multicolumn{4}{c}{FHI-aims} \\ \hline
HF(aimsRT2) & -1938.59753  &-1938.43178  & -1718.494936  & -1718.366256 \\
HF(aimsRT3+2) & -1938.63999 &-1938.46939  & -1718.520974 & -1718.391508 \\
B3LYP(aimsRT2) &-1943.40144  &-1943.44917  & -1722.185538 & -1722.13375 \\
B3LYP(aimsRT3+2) &-1943.406195  &-1943.453468  & -1722.188232 & -1722.136102 \\ \hline
\multicolumn{1}{c}{} & \multicolumn{4}{c}{GAMESS (cc-pVTZ)} \\ \hline
HF(gam) &  & -1938.589781 & -1938.425386 &  \\
B3LYP(gam) & -1942.914743 & -1942.965171 & -1721.803708 & -1721.754757 \\ \hline
 & \multicolumn{4}{c}{DMC} \\ \hline
DMC(B3LYP) & -1943.323152 & -1943.344858 & -1722.234695 & -1722.169160 \\
Error & 0.000713 & 0.000669 & 0.000546 & 0.000598 \\ \hline
\end{tabular}
\end{table}

\begin{table}[]
\centering
\caption{Raw data of [Fe(L)$_6$]$^{2+}$ where L is NCH and NH$_3$.}
\label{my-label}
\begin{tabular}{lcccc}
\hline
 & \multicolumn{2}{c}{NCH} & \multicolumn{2}{c}{NH$_3$} \\
 & HS & LS & HS & LS \\ \hline
\multicolumn{5}{c}{AB INITIO} \\ \hline
HF & -1819.620374 & -1819.487347 & -1599.491142 & -1599.371166 \\
CCSD & -1822.423771 & -1822.390926 & -1601.621124 & -1601.58094 \\
CCSD(T) & -1822.547794 & -1822.531749 & -1601.685784 & -1601.657385 \\ \hline
\multicolumn{5}{c}{SC-DFT} \\ \hline
SVWN & -1816.857548 & -1816.93601 & -1597.590343 & -1597.622558 \\
BP86 & -1824.320552 & -1824.353601 & -1603.094831 & -1603.095664 \\
BLYP & -1824.044252 & -1824.072593 & -1602.780882 & -1602.782754 \\
PBE & -1823.179736 & -1823.212964 & -1602.195969 & -1602.196691 \\
TPSS & -1824.237428 & -1824.271398 & -1602.943456 & -1602.948892 \\
TPSSh & -1824.142661 & -1824.15655 & -1602.892938 & -1602.884685 \\
B3LYP & -1823.74587 & -1823.739341 & -1602.608817 & -1602.58767 \\
PBE0 & -1823.202688 & -1823.187878 & -1602.255415 & -1602.224025 \\
HH & -1823.765913 & -1823.719894 & -1602.643047 & -1602.594119 \\ \hline
\multicolumn{5}{c}{HF-DFT} \\ \hline
SVWN(HF) & -1816.63952 & -1816.638734 & -1597.442603 & -1597.420226 \\
BP86(HF) & -1824.171878 & -1824.136674 & -1602.995131 & -1602.950198 \\
BLYP(HF) & -1823.875086 & -1823.842955 & -1602.665414 & -1602.62815 \\
PBE(HF) & -1823.032117 & -1822.995094 & -1602.099135 & -1602.052967 \\
TPSS(HF) & -1824.121705 & -1824.09658 & -1602.865679 & -1602.831769 \\
TPSSh(HF) & -1824.051686 & -1824.02009 & -1602.832908 & -1602.793536 \\
B3LYP(HF) & -1823.639404 & -1823.598725 & -1602.537834 & -1602.49317 \\
PBE0(HF) & -1823.125538 & -1823.076015 & -1602.206621 & -1602.150437 \\
HH(HF) & -1823.723703 & -1823.667414 & -1602.615548 & -1602.559095
\end{tabular}
\end{table}

\begin{table}[]
\centering
\caption{Raw data of [Fe(L)$_6$]$^{2+}$ where L is NCH and NH$_3$.}
\label{my-label}
\begin{tabular}{lcccc}
\hline
 & \multicolumn{2}{c}{NCH} & \multicolumn{2}{c}{NH$_3$} \\
 & HS & LS & HS & LS \\ \hline
\multicolumn{1}{c}{} & \multicolumn{4}{c}{GAMES/ECP} \\ \hline
B3LYP(gam) & -1823.742016 & -1823.738646 & -1602.604559 & -1602.586481 \\ \hline
 & \multicolumn{4}{c}{DMC} \\ \hline
DMC(B3LYP) & -1824.201405 & -1824.158306 & -1603.025729 & -1602.980673 \\
Error & 0.000968  & 0.000738  & 0.000565  & 0.000773  \\ \hline
\end{tabular}
\end{table}

\newpage
\begin{table}[]
\centering
\caption{[Fe(CO)$_6$]$^{2+}$ xyz data.}
\label{my-label}
\begin{tabular}{lllllllll}
\hline
CO\_HS &  &  &  &  & CO\_LS &  &  &  \\ \hline
13 &  &  &  &  & 13 &  &  &  \\
 &  &  &  &  &  &  &  &  \\
Fe & 0.0000013 & -0.0000001 & -0.0000001 &  & Fe & 0.0000002 & 0.0000002 & 0.0000001 \\
C & 0.0000037 & -0.0072194 & 2.3405884 &  & C & 0.0000001 & 0.0000002 & 1.9494602 \\
O & -0.0000299 & -0.0415619 & 3.4549199 &  & O & -0.0000002 & -0.0000001 & 3.0656647 \\
C & 2.3029923 & -0.0000137 & -0.0000135 &  & C & 1.9494601 & 0.0000000 & -0.0000001 \\
O & 3.4182461 & -0.0000170 & -0.0000168 &  & O & 3.0656646 & -0.0000003 & -0.0000005 \\
C & -2.3029877 & 0.0000133 & 0.0000131 &  & C & -1.9494597 & 0.0000000 & -0.0000001 \\
O & -3.4182415 & 0.0000167 & 0.0000165 &  & O & -3.0656642 & -0.0000003 & -0.0000005 \\
C & -0.0000019 & 0.0072196 & -2.3405883 &  & C & 0.0000001 & -0.0000001 & -1.9494600 \\
O & 0.0000250 & 0.0415622 & -3.4549197 &  & O & -0.0000003 & -0.0000004 & -3.0656645 \\
C & -0.0000028 & -2.3369861 & 0.0065791 &  & C & 0.0000000 & -1.9494597 & 0.0000002 \\
O & 0.0000240 & -3.4513475 & 0.0412328 &  & O & -0.0000003 & -3.0656642 & -0.0000001 \\
C & 0.0000037 & 2.3369862 & -0.0065788 &  & C & 0.0000000 & 1.9494600 & 0.0000001 \\
O & -0.0000326 & 3.4513476 & -0.0412324 &  & O & -0.0000003 & 3.0656646 & 0.0000004
\end{tabular}
\end{table}

\newpage
\begin{table}[]
\centering
\caption{[Fe(NCH)$_6$]$^{2+}$ xyz data.}
\label{my-label}
\begin{tabular}{lllllllll}
\hline
NCH\_HS &  &  &  &  & NCH\_LS &  &  &  \\ \hline
19 &  &  &  &  & 19 &  &  &  \\
 &  &  &  &  &  &  &  &  \\
Fe & 0.0000000 & 0.0000005 & -0.0000008 &  & Fe & 0.0000000 & 0.0000000 & 0.0000003 \\
N & 0.0000002 & 2.2062681 & -0.0000002 &  & N & 0.0000000 & 1.9756648 & 0.0000002 \\
N & -2.2009845 & 0.0000002 & -0.0000003 &  & N & -1.9756648 & 0.0000000 & 0.0000002 \\
N & -0.0000004 & -0.0000004 & -2.2009794 &  & N & -0.0000001 & 0.0000001 & -1.9756664 \\
N & -0.0000004 & -0.0000004 & 2.2009781 &  & N & 0.0000000 & 0.0000001 & 1.9756669 \\
N & 2.2009844 & 0.0000010 & -0.0000003 &  & N & 1.9756648 & -0.0000001 & 0.0000002 \\
N & 0.0000007 & -2.2062673 & -0.0000003 &  & N & 0.0000000 & -1.9756648 & 0.0000002 \\
C & 0.0000012 & -3.3493502 & 0.0000003 &  & C & 0.0000004 & -3.1177203 & -0.0000002 \\
H & 0.0000016 & -4.4237178 & 0.0000007 &  & H & 0.0000007 & -4.1914191 & -0.0000004 \\
C & -3.3446088 & 0.0000009 & 0.0000007 &  & C & -3.1177203 & -0.0000005 & -0.0000002 \\
H & -4.4184325 & 0.0000014 & 0.0000013 &  & H & -4.1914191 & -0.0000008 & -0.0000005 \\
C & -0.0000010 & -0.0000013 & -3.3446039 &  & C & -0.0000004 & 0.0000004 & -3.1177221 \\
H & -0.0000013 & -0.0000020 & -4.4184276 &  & H & -0.0000007 & 0.0000007 & -4.1914211 \\
C & 3.3446087 & 0.0000005 & 0.0000006 &  & C & 3.1177203 & -0.0000005 & -0.0000002 \\
H & 4.4184324 & 0.0000002 & 0.0000013 &  & H & 4.1914191 & -0.0000008 & -0.0000005 \\
C & -0.0000010 & -0.0000013 & 3.3446025 &  & C & -0.0000005 & 0.0000004 & 3.1177226 \\
H & -0.0000014 & -0.0000020 & 4.4184262 &  & H & -0.0000008 & 0.0000007 & 4.1914216 \\
C & 0.0000007 & 3.3493510 & 0.0000003 &  & C & 0.0000004 & 3.1177203 & -0.0000002 \\
H & 0.0000011 & 4.4237186 & 0.0000007 &  & H & 0.0000007 & 4.1914191 & -0.0000004
\end{tabular}
\end{table}

\newpage
\begin{table}[]
\centering
\caption{[Fe(NH$_3$)$_6$]$^{2+}$ xyz data.}
\label{my-label}
\begin{tabular}{lllllllll}
\hline
NH3\_HS &  &  &  &  & NH3\_LS &  &  &  \\ \hline
25 &  &  &  &  & 25 &  &  &  \\
 &  &  &  &  &  &  &  &  \\
Fe & -1.6827629 & -0.5620638 & 0.0006858 &  & Fe & -1.6825128 & -0.5610153 & -0.0006961 \\
N & -1.7654701 & 1.7386430 & -0.0155915 &  & N & -1.7424118 & 1.5552120 & 0.0158045 \\
H & -2.3091690 & 2.1062136 & 0.7643967 &  & H & -2.4036875 & 1.9339858 & 0.6938915 \\
H & -0.8575605 & 2.1961896 & 0.0569999 &  & H & -0.8559216 & 2.0036953 & 0.2471666 \\
H & -2.1985074 & 2.1444062 & -0.8446345 &  & H & -2.0193966 & 1.9820701 & -0.8682401 \\
N & 0.6370910 & -0.4829211 & 0.0198374 &  & N & 0.4336893 & -0.5015161 & 0.0181596 \\
H & 1.0410909 & -0.1381987 & 0.8898340 &  & H & 0.8440623 & -0.2468840 & 0.9166699 \\
H & 1.0933299 & -1.3794527 & -0.1444124 &  & H & 0.8887488 & -1.3803696 & -0.2288050 \\
H & 1.0148101 & 0.1305669 & -0.7014791 &  & H & 0.8222226 & 0.1746500 & -0.6393939 \\
N & -4.0024708 & -0.5711646 & -0.0821798 &  & N & -3.7986910 & -0.5778442 & -0.0617939 \\
H & -4.4120209 & -1.4547568 & -0.3829660 &  & H & -4.2102994 & -1.4780144 & -0.3084425 \\
H & -4.4588691 & -0.3582133 & 0.8039661 &  & H & -4.2539588 & -0.3220773 & 0.8144309 \\
H & -4.3743783 & 0.1202099 & -0.7325748 &  & H & -4.1858156 & 0.0740192 & -0.7442454 \\
N & -1.6769265 & -2.8574940 & 0.0683045 &  & N & -1.6610569 & -2.6779301 & 0.0242933 \\
H & -2.4248044 & -3.2491480 & 0.6399273 &  & H & -2.3903796 & -3.0814636 & 0.6123051 \\
H & -1.8046273 & -3.2785580 & -0.8511997 &  & H & -1.7968348 & -3.1174279 & -0.8861569 \\
H & -0.8224786 & -3.2773553 & 0.4326991 &  & H & -0.7980809 & -3.0901543 & 0.3789793 \\
N & -1.6947644 & -0.6233020 & 2.2950899 &  & N & -1.7053421 & -0.5825425 & 2.1162469 \\
H & -0.9567299 & -1.2059553 & 2.6891561 &  & H & -0.9813845 & -1.1763493 & 2.5209486 \\
H & -1.5535805 & 0.2946430 & 2.7152532 &  & H & -1.5616171 & 0.3273586 & 2.5543575 \\
H & -2.5558826 & -0.9737822 & 2.7131764 &  & H & -2.5715855 & -0.9289025 & 2.5286917 \\
N & -1.5929695 & -0.5620411 & -2.2999504 &  & N & -1.6210680 & -0.5813637 & -2.1168366 \\
H & -1.0185358 & -1.3236265 & -2.6592866 &  & H & -0.9545040 & -1.2553296 & -2.4936978 \\
H & -2.4963312 & -0.6738135 & -2.7585987 &  & H & -2.5055538 & -0.8204390 & -2.5652371 \\
H & -1.1916622 & 0.2796750 & -2.7122128 &  & H & -1.3505210 & 0.3037723 & -2.5455600
\end{tabular}
\end{table}

\newpage
\begin{table}[]
\centering
\caption{[Fe(H$_2$O)$_6$]$^{2+}$ xyz data.}
\label{my-label}
\begin{tabular}{lllllllll}
\hline
H2O\_HS &  &  &  &  & H2O\_LS &  &  &  \\ \hline
19 &  &  &  &  & 19 &  &  &  \\
 &  &  &  &  &  &  &  &  \\
Fe & -1.6845392 & -0.7206701 & -0.1386922 &  & Fe & -1.6845545 & -0.7208248 & -0.1387753 \\
O & -1.7225165 & 1.4435451 & -0.0788233 &  & O & -1.8371609 & 1.3103074 & -0.2366417 \\
H & -2.2327347 & 2.0721657 & -0.6081837 &  & H & -2.7248031 & 1.6951387 & -0.1781363 \\
H & -1.1836565 & 1.9689883 & 0.5299181 &  & H & -1.2260721 & 1.9506903 & 0.1550818 \\
O & 0.4742655 & -0.5541733 & -0.1258573 &  & O & 0.3478431 & -0.6860383 & 0.0239942 \\
H & 1.1321101 & -1.0258989 & 0.4036902 &  & H & 0.8167239 & -1.5311871 & -0.0487537 \\
H & 0.9645036 & 0.0562592 & -0.6951254 &  & H & 0.9394100 & -0.0045677 & -0.3261609 \\
O & -3.8432628 & -0.8879805 & -0.1526380 &  & O & -3.7191775 & -0.7293063 & -0.2749855 \\
H & -4.3335551 & -1.4967341 & 0.4183794 &  & H & -4.3343364 & -1.1264790 & 0.3580976 \\
H & -4.5012426 & -0.4156202 & -0.6814497 &  & H & -4.1078491 & -0.8432878 & -1.1555310 \\
O & -1.6468488 & -2.8848921 & -0.1975952 &  & O & -1.5049607 & -2.7508363 & -0.0662374 \\
H & -2.1866691 & -3.4101261 & -0.8056683 &  & H & -1.8213499 & -3.3785131 & -0.7317227 \\
H & -1.1337667 & -3.5136153 & 0.3288629 &  & H & -1.6098710 & -3.1783680 & 0.7972597 \\
O & -1.8336924 & -0.6522882 & 2.0203864 &  & O & -1.6955636 & -0.7997349 & 1.8988762 \\
H & -1.3324137 & -1.1313887 & 2.6948456 &  & H & -0.8512825 & -0.6664715 & 2.3558853 \\
H & -2.4655514 & -0.0913583 & 2.4927695 &  & H & -2.3885668 & -0.4582020 & 2.4820276 \\
O & -1.5352491 & -0.7884118 & -2.2977680 &  & O & -1.6983650 & -0.6692265 & -2.1773196 \\
H & -0.9010307 & -1.3468668 & -2.7699235 &  & H & -1.2754814 & -1.3071697 & -2.7698947 \\
H & -2.0404799 & -0.3138537 & -2.9724876 &  & H & -1.6308724 & 0.1989462 & -2.6029238 \\
\end{tabular}
\end{table}

\begin{table}[]
\centering
\caption{$Fe^{2+}$ atomic charge of [Fe(L)$_6$]$^{2+}$ with various Metal-Ligand distance. DFT shows insufficient atomic charge due to the self-interaction error while HF does not.}
\label{my-label}
\begin{tabular}{lcccclccccc}
\hline
 & RM-L & HF & PBE &  &  & LS &  & HF & PBE & \\ \hline
CO\_HS & Req=2.3 & 1.08 & 0.57 &  &  & CO\_LS & Req=1.9 & 0.33 & -0.39 &  \\
 & Rinf=5.0 & 1.94 & 1.16 &  &  &  & Rinf=5.0 & 1.93 & 1.05 &  \\
NCH\_HS & Req=2.2 & 1.70 & 1.04 &  &  & NCH\_LS & Req=2.0 & 1.37 & 0.56 &  \\
 & Rinf=5.0 & 1.95 & 1.21 &  &  &  & Rinf=5.0 & 1.95 & 1.13 &  \\
NH3\_HS & Req=2.3 & 1.46 & 0.72 &  &  & NH3\_LS & Req=2.1 & 1.32 & 0.28 &  \\
 &  &  &  &  &  &  & Rinf=5.0 &  & 0.76 &  \\
H2O\_HS & Req=2.2 & 1.43 & 0.79 &  &  & H2O\_LS & Req=2.0 & 1.31 & 0.51 &  \\
 & Rinf=5.0 & 1.95 & 1.02 &  &  &  & Rinf=5.0 & 1.94 & 0.92 & 
\end{tabular}
\end{table}

\begin{table}[]
\centering
\caption{[Fe(L)$_6$]$^{2+}$ HOMO-LUMO gap of HF and PBE. Note that all units are eV.}
\label{my-label}
\begin{tabular}{lcc}
\hline
 & HF & PBE \\ \hline
CO\_HS	 & 15.33 & 0.45 \\
CO\_LS 	& 17.99 & 3.98 \\
NCH\_HS	 & 15.04 & 0.43 \\
NCH\_LS	 & 16.38 & 2.44 \\
NH3\_HS 	& 13.82 & 0.47 \\
NH3\_LS	 & 15.40 & 1.91 \\
H2O\_HS	 & 14.57 & 0.66 \\
H2O\_LS	 & 16.25 & 1.38
\end{tabular}
\end{table}

\clearpage
\section{Data with TZVP basis set} 

\begin{table}[]
\centering
\caption{Raw data of [Fe(L)$_6$]$^{2+}$ where L is CO, and H$_2$O, TZVP basis set.}
\label{my-label}
\begin{tabular}{lcccc}
\hline
 & \multicolumn{2}{c}{CO} & \multicolumn{2}{c}{H2O} \\
 & HS & LS & HS & LS \\ \hline
\multicolumn{5}{c}{AB INITIO} \\ \hline
HF & -1938.544195 & -1938.368443 & -1718.45312 & -1718.315537 \\
CCSD & -1940.931932 & -1940.919423 & -1720.182227 & -1720.098026 \\
CCSD(T) & -1941.026375 & -1941.038878 & -1720.220097 & -1720.141748 \\ \hline
\multicolumn{5}{c}{SC-DFT} \\ \hline
SVWN\_TZVP & -1935.562476 & -1935.75007 & -1716.324658 & -1716.307699 \\
BP86\_TZVP & -1943.490062 & -1943.609566 & -1722.272681 & -1722.230268 \\
BLYP\_TZVP & -1943.245756 & -1943.345428 & -1722.004401 & -1721.967074 \\
PBE\_TZVP & -1942.28028 & -1942.403351 & -1721.304821 & -1721.261134 \\
TPSS\_TZVP & -1943.39643 & -1943.507391 & -1722.107634 & -1722.072356 \\
TPSSh\_TZVP & -1943.282565 & -1943.36364 & -1722.043473 & -1721.998094 \\
B3LYP\_TZVP & -1942.879188 & -1942.924235 & -1721.775629 & -1721.721931 \\
PBE0\_TZVP & -1942.269433 & -1942.316757 & -1721.343075 & -1721.276812 \\
HH\_TZVP & -1942.883119 & -1942.853297 & -1721.805249 & -1721.730594 \\ \hline
\multicolumn{5}{c}{HF-DFT} \\ \hline
SVWN(HF)\_TZVP & -1935.299796 & -1935.378935 & -1716.15509 & -1716.105853 \\
BP86(HF)\_TZVP & -1943.333361 & -1943.356739 & -1722.1811 & -1722.114554 \\
BLYP(HF)\_TZVP & -1943.071378 & -1943.084017 & -1721.897645 & -1721.840542 \\
PBE(HF)\_TZVP & -1942.122074 & -1942.145724 & -1721.215951 & -1721.147715 \\
TPSS(HF)\_TZVP & -1943.270719 & -1943.296682 & -1722.034244 & -1721.977504 \\
TPSSh(HF)\_TZVP & -1943.184032 & -1943.197198 & -1721.987336 & -1721.924908 \\
B3LYP(HF)\_TZVP & -1942.771561 & -1942.763296 & -1721.711073 & -1721.645457 \\
PBE0(HF)\_TZVP & -1942.187458 & -1942.180504 & -1721.299684 & -1721.220548 \\
HH(HF)\_TZVP & -1942.841231 & -1942.79396 & -1721.780398 & -1721.702109
\end{tabular}
\end{table}

\newpage
\begin{table}[]
\centering
\caption{Raw data of [Fe(L)$_6$]$^{2+}$ where L is NCH and NH$_3$, TZVP basis set.}
\label{my-label}
\begin{tabular}{lcccc}
\hline
 & \multicolumn{2}{c}{NCH} & \multicolumn{2}{c}{NH$_3$} \\
 & HS & LS & HS & LS \\ \hline
\multicolumn{5}{c}{AB INITIO} \\ \hline
HF & -1819.575158 & -1819.435748 & -1599.452665 & -1599.327687 \\
CCSD & -1821.896763 & -1821.846661 & -1601.123217 & -1601.069463 \\
CCSD(T) & -1821.990379 & -1821.952622 & -1601.165067 & -1601.119688 \\ \hline
\multicolumn{5}{c}{SC-DFT} \\ \hline
SVWN\_TZVP & -1816.857548 & -1816.93601 & -1597.590343 & -1597.622558 \\
BP86\_TZVP & -1824.284226 & -1824.315432 & -1603.057978 & -1603.056034 \\
BLYP\_TZVP & -1824.005452 & -1824.032018 & -1602.741823 & -1602.741127 \\
PBE\_TZVP & -1823.14398 & -1823.175345 & -1602.159974 & -1602.157843 \\
TPSS\_TZVP & -1824.201308 & -1824.234426 & -1602.909479 & -1602.913085 \\
TPSSh\_TZVP & -1824.106194 & -1824.119101 & -1602.85921 & -1602.849486 \\
B3LYP\_TZVP & -1823.706825 & -1823.698327 & -1602.571109 & -1602.548127 \\
PBE0\_TZVP & -1823.166204 & -1823.149511 & -1602.2208 & -1602.187785 \\
HH\_TZVP & -1823.726392 & -1823.67743 & -1602.606373 & -1602.555522 \\ \hline
\multicolumn{5}{c}{HF-DFT} \\ \hline
SVWN(HF)\_TZVP & -1816.613916 & -1816.618223 & -1597.415187 & -1597.399208 \\
BP86(HF)\_TZVP & -1824.143582 & -1824.112787 & -1602.966054 & -1602.926874 \\
BLYP(HF)\_TZVP & -1823.846515 & -1823.817712 & -1602.635138 & -1602.602517 \\
PBE(HF)\_TZVP & -1823.003816 & -1822.971334 & -1602.070621 & -1602.03038 \\
TPSS(HF)\_TZVP & -1824.090705 & -1824.069565 & -1602.83662 & -1602.80771 \\
TPSSh(HF)\_TZVP & -1824.019271 & -1823.990722 & -1602.80304 & -1602.767774 \\
B3LYP(HF)\_TZVP & -1823.607715 & -1823.568499 & -1602.50611 & -1602.464246 \\
PBE0(HF)\_TZVP & -1823.093083 & -1823.045631 & -1602.176074 & -1602.123345 \\
HH(HF)\_TZVP & -1823.687019 & -1823.629106 & -1602.5811 & -1602.524398
\end{tabular}
\end{table}

\newpage

\begin{table}[]
\centering
\caption{Raw data of [Fe(L)$_6$]$^{2+}$, cc-pVDZ basis where L is CO, H$_2$O, NCH and NH$_3$}
\label{my-label}
\begin{tabular}{lcc}
\hline
        & \multicolumn{2}{c}{cc-pVDZ} \\
        & CCSD(T)      & CCSD         \\
\hline
CO\_HS  & -1940.683472 & -1940.601943 \\
CO\_LS  & -1940.718958 & -1940.608786 \\
H2O\_HS & -1720.005539 & -1719.974971 \\
H2O\_LS & -1719.933587 & -1719.896417 \\
NCH\_HS & -1821.683798 & -1821.600966 \\
NCH\_LS & -1821.655531 & -1821.559138 \\
NH3\_HS & -1601.02068  & -1600.983348 \\
NH3\_LS & -1600.984901 & -1600.937788
\end{tabular}
\end{table}

\begin{table}[]
\centering
\caption{[Fe(L)$_6$]$^{2+}$ D1 diagnostics (CCSD) of cc-pVDZ, cc-pVTZ and TZVP basis sets.}
\label{my-label}
\begin{tabular}{lccc}
\hline
        & cc-pVDZ & cc-pVTZ & TZVP   \\
\hline
CO\_HS  & 0.0598  & 0.0602  & 0.0538 \\
CO\_LS  & 0.1419  & 0.1438  & 0.1305 \\
H2O\_HS & 0.0374  & 0.0381  & 0.0339 \\
H2O\_LS & 0.059   & 0.0604  & 0.0536 \\
NCH\_HS & 0.0591  & 0.0508  & 0.0503 \\
NCH\_LS & 0.0866  & 0.0874  & 0.0853 \\
NH3\_HS & 0.0562  & 0.0575  & 0.0494 \\
NH3\_LS & 0.0865  & 0.0872  & 0.076 
\end{tabular}
\end{table}
\clearpage
\section{List of Figures}

\noindent{Figure 1. Spin adiabatic energy differences in eV of Fe complexes for various DFT calculations and CCSD(T). All DFT and HF-DFT used TZVP basis set.}\\

\noindent{Figure 2. SA rainbow plot of NCH complex evaluated with several different XC approximations 
on different self-consistent densities (red dots) and the HF density (blue bar) using TZVP basis set. The x-axis labels
the energy functional, the colored bars indicate which density (grey is LDA, yellow
is GGA/mGGA, green is hybrid, and blue is HF).  
The dark green bar uses the self-consistent density of Becke's half-and-half functional (HH),
which contains 50\% exact exchange.}\\

\noindent{Figure 3. $a$PBE and HF-$a$PBE of [Fe(NCH)$_6$]$^{2+}$, TZVP basis set.}

\begin{figure}[htb]
\includegraphics[width=0.8\columnwidth]{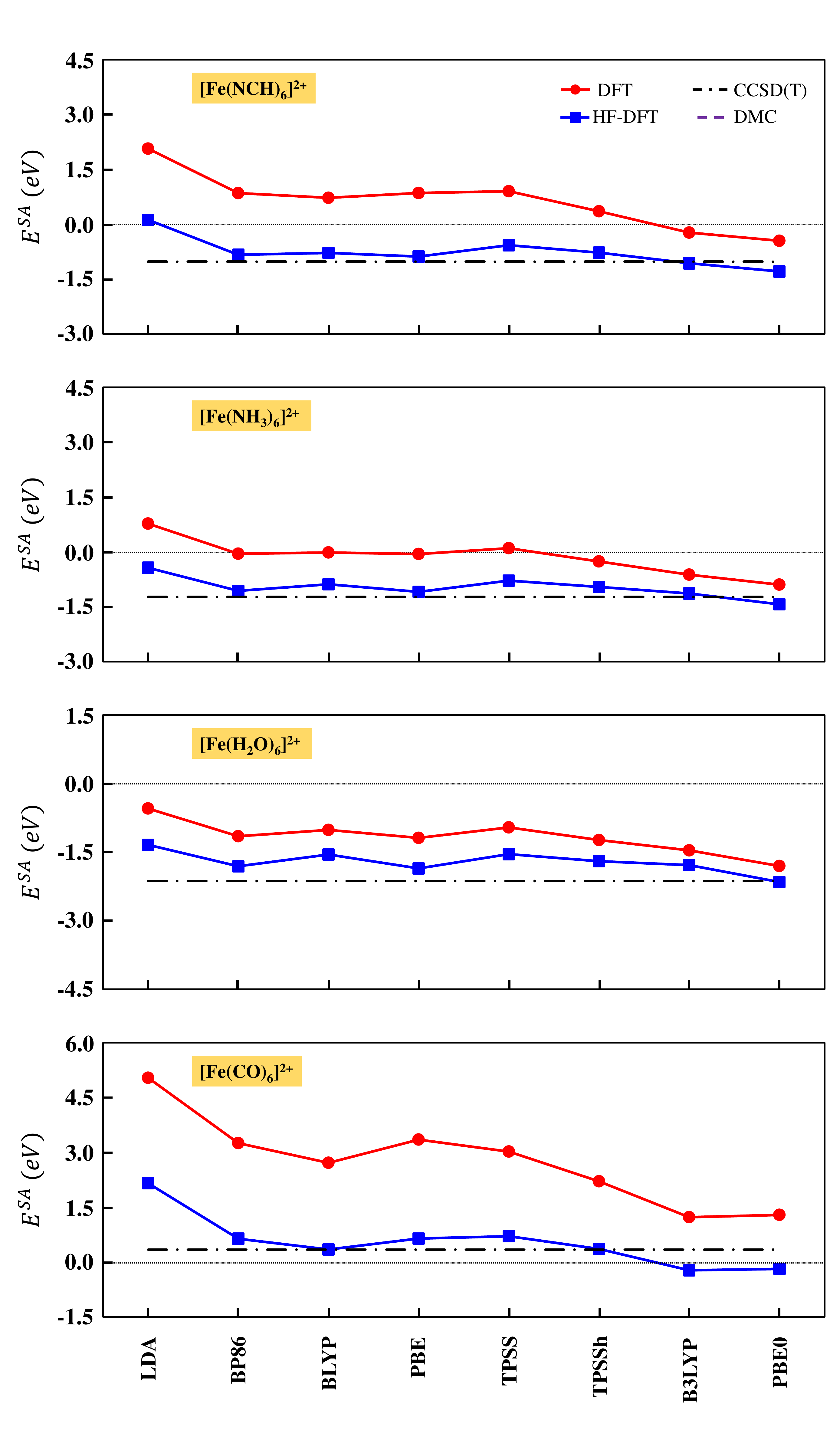}
\caption{Spin adiabatic energy differences in eV of Fe complexes for various DFT calculations and CCSD(T). All DFT and HF-DFT used TZVP basis set.}
\label{SuperFigure}
\end{figure}

\begin{figure}[htb]
\includegraphics[width=1.0\columnwidth]{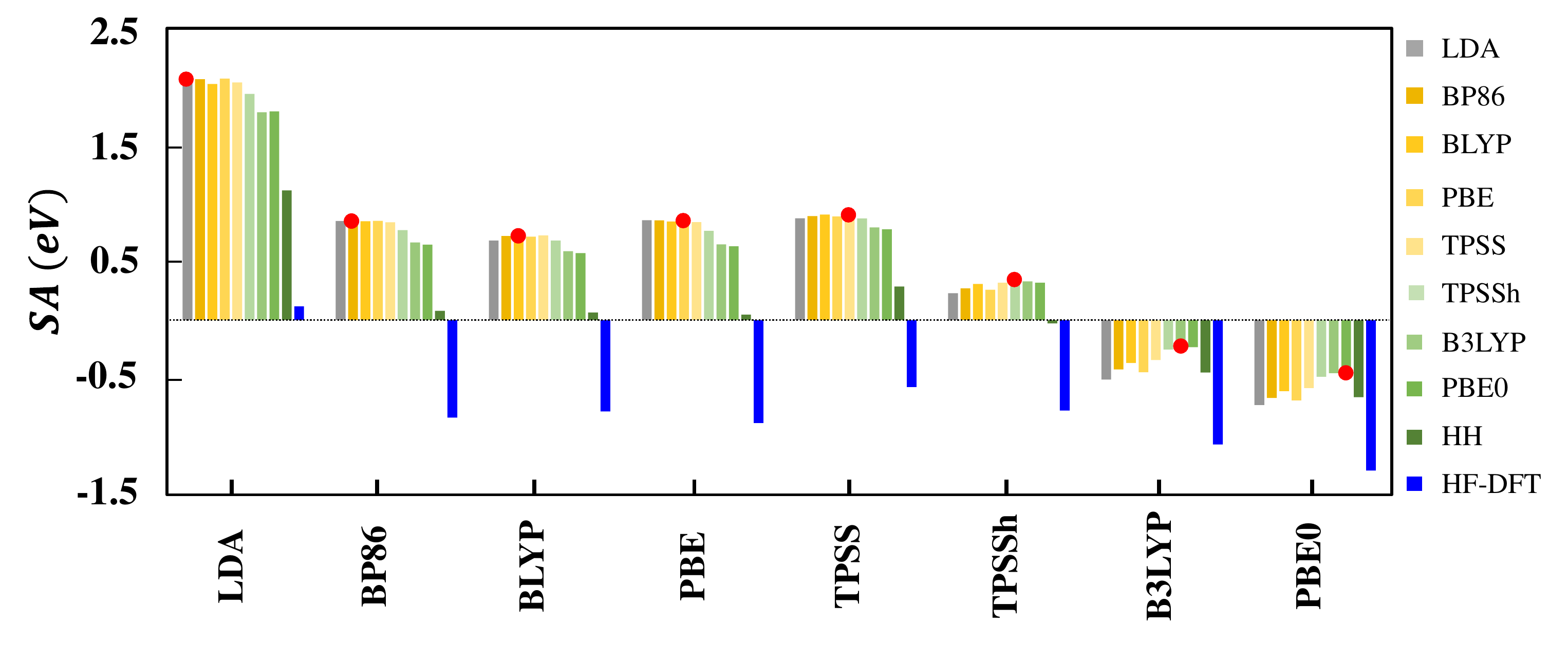}
\caption{SA rainbow plot of NCH complex evaluated with several different XC approximations 
on different self-consistent densities (red dots) and the HF density (blue bar) using TZVP basis set. The x-axis labels
the energy functional, the colored bars indicate which density (grey is LDA, yellow
is GGA/mGGA, green is hybrid, and blue is HF).  
The dark green bar uses the self-consistent density of Becke's half-and-half functional (HH),
which contains 50\% exact exchange.}
\label{rainbow}
\end{figure}

\begin{figure}[htb]
\includegraphics[width=0.8\columnwidth]{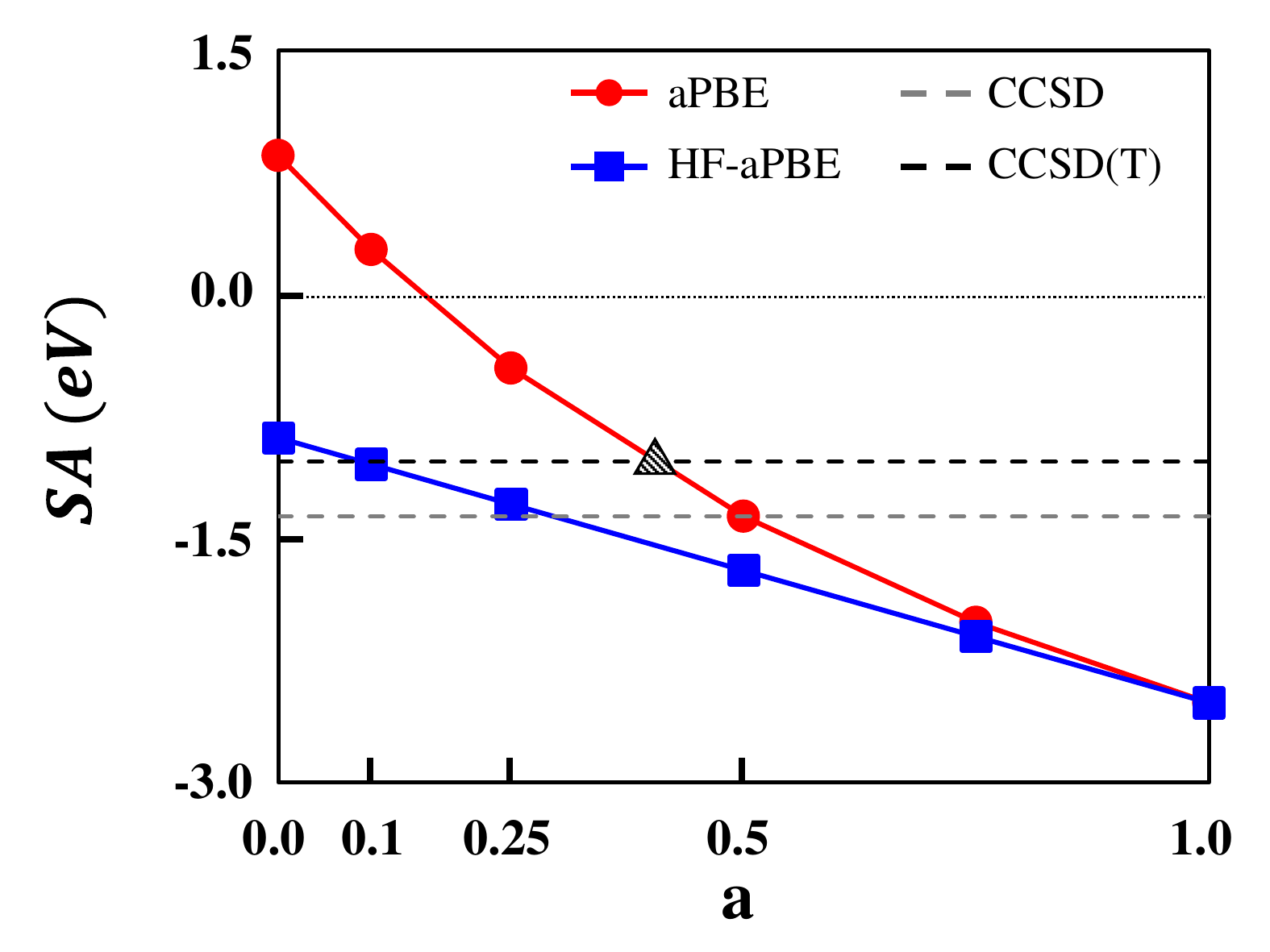}
\caption{$a$PBE and HF-$a$PBE of [Fe(NCH)$_6$]$^{2+}$, TZVP basis set.}
\label{fig:dE}
\end{figure}


\clearpage
\providecommand{\latin}[1]{#1}
\makeatletter
\providecommand{\doi}
  {\begingroup\let\do\@makeother\dospecials
  \catcode`\{=1 \catcode`\}=2 \doi@aux}
\providecommand{\doi@aux}[1]{\endgroup\texttt{#1}}
\makeatother
\providecommand*\mcitethebibliography{\thebibliography}
\csname @ifundefined\endcsname{endmcitethebibliography}
  {\let\endmcitethebibliography\endthebibliography}{}

\end{document}